\documentclass[preprint,superscriptaddress,aps]{revtex4}
\usepackage{amsfonts}
\usepackage{graphics}
\usepackage{graphicx}
\usepackage{slashed}
\newcommand{\be}{\begin{equation}}
\newcommand{\ee}{\end{equation}}
\newcommand{\en}{\end{equation}}
\newcommand{\ba}{\begin{eqnarray}}
\newcommand{\ea}{\end{eqnarray}}
\newcommand{\bea}{\begin{eqnarray}}
\newcommand{\eea}{\end{eqnarray}}
\newcommand{\bq}{\begin{eqnarray}}
\newcommand{\eq}{\end{eqnarray}}

\def\pls{\partial\!\!\!/}

\def\As{A\!\!\!/}

\def\ls{l\!\!\!/}
\def\ps{p\!\!\!/}

\def\bs{b\!\!\!/}

\def\ds{\partial\!\!\!/}

\begin{document}

\title{Four-dimensional aether-like Lorentz-breaking QED revisited and problem of ambiguities}
\author{A. P. Baeta Scarpelli}
\affiliation{Setor T\'{e}cnico-Cient\'{\i}fico - Departamento de Pol\'\i cia Federal
Rua Hugo D'Antola, 95 - Lapa - S\~{a}o Paulo - Brazil}
\email{scarpelli.apbs@dpf.gov.br}
\author{T. Mariz}
\affiliation{Instituto de F\'\i sica, Universidade Federal de Alagoas, 57072-270, Macei\'o, Alagoas, Brazil}
\email{tmariz@fis.ufal.br}
\author{J. R. Nascimento}
\affiliation{Departamento de F\'{\i}sica, Universidade Federal da Para\'{\i}ba\\
 Caixa Postal 5008, 58051-970, Jo\~ao Pessoa, Para\'{\i}ba, Brazil}
\email{jroberto, petrov@fisica.ufpb.br}
\author{A. Yu. Petrov}
\affiliation{Departamento de F\'{\i}sica, Universidade Federal da Para\'{\i}ba\\
 Caixa Postal 5008, 58051-970, Jo\~ao Pessoa, Para\'{\i}ba, Brazil}
\email{jroberto, petrov@fisica.ufpb.br}
\begin{abstract}
In this paper, we consider the perturbative generation of the CPT-even aether-like Lorentz-breaking term in the extended Lorentz-breaking QED within different approaches and discuss its ambiguities.
\end{abstract}

\maketitle

\section{Introduction}
\label{intro}
The interest in the study of different aspects of Lorentz symmetry breaking is very high now. Initially, it has been motivated by studies in string theory which have shown that the Lorentz symmetry breaking can emerge naturally when the perturbative string vacuum is unstable, since in this case some tensors naturally acquire non-zero vacuum expectations introducing thus preferred directions in the spacetime \cite{KostSam,KostSam1}.

The first known example of a Lorentz-breaking extension for field theory models, that is, the electrodynamics with the Lorentz breaking Carroll-Field-Jackiw (CFJ) term, has been introduced in \cite{CFJ}. This term, which can be radiatively induced if a Lorentz-violating axial term $\bar{\psi}\bs \gamma_5\psi$ is included in the fermionic sector, was shown to break also the CPT symmetry \cite{Kostcoll,Kostcoll1}. Different results for the CFJ term have been obtained in a number of papers, see f.e. \cite{list,list1,list2}. The key property of this term consists in its ambiguity \cite{Kostcoll,Kostcoll1}, i.e. this term, arisen as a quantum correction, is finite, but the result for it depends on the regularization scheme, since the corresponding contribution is superficially divergent.  This ambiguity turns out to be related to the axial anomaly \cite{JackAmb}, and it was shown in \cite{Chung} that the effect of the completely undetermined value of the CFJ coefficient naturally emerges within the functional integral formalism. An extensive discussion of this ambiguity is presented in \cite{Alt1,Alt2}. Further, this analogues of the CFJ coefficient were shown to be undetermined also in the non-Abelian extension of the Lorentz-breaking QED \cite{ourLV1}, and in the finite temperature case \cite{ourLV1,ourLV2}.

At the same time, the CFJ term is not the unique term displaying the ambiguity in the one-loop approximation. A similar situation takes place also for the four-dimensional gravitational Chern-Simons term \cite{JaPi} whose perturbative generation has been discussed in details in \cite{ourgra}.

However, both CFJ and gravitational Chern-Simons terms break not only the Lorentz symmetry but also the CPT symmetry. Therefore, the natural question is whether the ambiguity can emerge for the CPT-even terms, those ones proportional to a constant even-rank tensor. Recently,  the four-dimensional aether-like term in the nonmimimal QED, with a magnetic coupling only, was discussed and shown to be ambiguous \cite{aether,aether1} (a number of issues related to this term has been discussed in \cite{Belich0,Belich1,Belich2,Belich3,Ferr,Ferr1,Ferr2,Ferr3}). Further, in \cite{MCN} it was shown that the ambiguity vanishes in a purely nonminimal sector, if we consider a gauge theory involving both minimal and nonminimal couplings, and impose a gauge-preserving regularization. However, the problem of ambiguity of the aether term is still open, even in the theory involving two couplings -- it is worth mentioning that the three-derivative Myers-Pospelov term was shown to be ambiguous being generated on the base of the theory with two couplings \cite{MNP}.

The aim of this paper consists in the one-loop perturbative generation of the aether-like term in four-dimensional Lorentz-breaking QED with two couplings and an axial term in the fermionic sector. We will show that the result for it is not exhausted by the contributions discussed in \cite{aether,MCN} since it involves new terms, and, moreover, displays the same ambiguity as the CFJ term.

\section{Generation of the aether term for the massive fermions}
\label{sec:1}

To begin our study, we formulate the extended spinor electrodynamics which involves two couplings (the minimal one, proportional to $e$, and the nonminimal one, proportional to $g$) and an axial term in the fermionic sector:
\begin{equation}
\label{mcn}
{\cal L}=\bar{\psi}\left[i \pls- \gamma^{\mu}(eA_{\mu}+g\epsilon_{\mu\nu\lambda\rho}F^{\nu\lambda}b^{\rho}) - m -  \gamma_{5}\bs\right]\psi.
\end{equation}
We are interested in presenting of a model in which a CPT-even term in the gauge
sector is not only induced, but also manifests an ambiguity in its coefficients. The uni\-que\-ness of this model is based on the fact that it, being a Lorentz-breaking extension of QED, is the only one allowing the emergence of finite Lorentz-breaking quantum correction while, for example, the models involving higher-derivative terms in the fermionic sector,  as well as the models involving the CPT-even Lorentz-breaking terms at the tree level, can be shown to yield divergent corrections requiring renormalization  (cf. the discussion in \cite{MNP}), therefore, in those cases such terms must be introduced in the theory from the very beginning whereas in our case they arise as consistent quantum corrections.

Unlike the model considered in \cite{aether}, this model can be reduced to the usual extended QED in the limit $g\to 0$. Namely this action has been considered in \cite{MCN,MNP}.
Here the $b_\rho$ is a constant vector implementing the Lorentz symmetry breaking, and $F_{\mu\nu}=\partial_{\mu}A_{\nu}-\partial_{\nu}A_{\mu}$ is the usual stress tensor constructed on the base of the gauge field $A_{\mu}$.

The importance of the action (\ref{mcn}) is motivated by the fact that it yields finite corrections to the gauge sector. Just this situation takes place for the higher-derivative contributions \cite{MNP}. Therefore, the model (\ref{mcn}) is the simplest Lorentz-CPT breaking model allowing for the finite aether-like quantum corrections, being at the same time a natural extension of the QED.

Let us now treat the gauge field as a purely external one, just as within the Schwinger approach. In this case the theory will be renormalizable. And its (one-loop) quantum correction in the effective action looks like
\begin{eqnarray}
S_{eff}[b,A]&=&-i\,{\rm Tr}\,\ln(i\partial\!\!\! /-e\gamma^{\mu}A_{\mu}-g\epsilon_{\mu\nu\lambda\rho}\gamma^{\mu}F^{\nu\lambda}b^{\rho}- %\nonumber\\&-&
m - \gamma_5 \bs),
\end{eqnarray}
The correction of the second order in the Lorentz- breaking vector $b_{\mu}$ in a purely nonminimal sector, where $e=0$, has been discussed in details in \cite{aether,MCN}. It was shown there that this correction looks like
\begin{eqnarray}
\label{s2d40}
&&S_{FF}(p)=-\frac{g^2}{2}\epsilon^{\alpha\beta\gamma\delta}\epsilon^{\alpha'\beta'\gamma'\delta'}
%\times\nonumber\\&\times&
b_{\alpha}F_{\beta\gamma}(p)
b_{\alpha'}
F_{\beta'\gamma'}(-p)\\&\times&
\int\frac{d^4k}{(2\pi)^4}\frac{1}{[k^2-m^2]^2}{\rm tr}
\big[m^2\gamma_\delta\gamma_{\delta'}+k^\mu k^\nu\gamma_\mu\gamma_\delta\gamma_\nu
\gamma_{\delta'}\big],\nonumber
\end{eqnarray}
which yields the result
\begin{eqnarray}
\label{s2d4}
\label{ff}
S_{FF}(p)&=&C_0g^2m^2(b^\alpha F_{\alpha\beta})^2,
\end{eqnarray}
where the constant $C_0$ is known to be equal either to $\frac{1}{4\pi^2}$ or to zero, see \cite{aether,aether1}. This is just the aether term proposed in \cite{Carroll}. It represents itself as a particular form of the most general CPT-even term $k^{\alpha\beta\gamma\delta}F_{\alpha\beta}F_{\gamma\delta}$ whose properties at the tree level have been studied in \cite{Ferr,Ferr1,Ferr2,Ferr3}.
Further, in \cite{MCN} it was shown that the value $C_0=0$ is preferable since the same constant arises during the calculation of the Proca-like correction, whose vanishing is natural from the viewpoint of the gauge invariance (note however that, in principle, it is not forbidden to use different values for the constant $C_0$ when Proca-like and aether-like terms are considered, since the constant $C_0$ is regularization dependent). Nevertheless, it is very important that within the model (\ref{mcn}) there exists a much more powerful source of ambiguities which we begin to discuss now.

So, let us turn to the aether-like corrections essentially depending on $e$. There are two such correction: one is purely minimal, proportional to $e^2$, and the other one is nonminimal, proportional to $eg$.

First, the correction of the second order in the Lorentz-breaking vector $b_{\mu}$ in a purely minimal sector, where $g=0$, is given by the following expression:
%\begin{widetext}
\begin{eqnarray}
\label{1}
S_{AA}(p)&=&\frac{ie^2}{2}\int\frac{d^4l}{(2\pi)^4}\mbox{tr}(\gamma^{\mu}\frac{1}{\ls-m}\gamma^{\nu}\frac{1}{\ls+\ps-m}
%\times\nonumber\\&\times&
\gamma_5\bs\frac{1}{\ls+\ps-m}\gamma_5\bs\frac{1}{\ls+\ps-m}+ 
\nonumber\\&+&
\gamma^{\mu}\frac{1}{\ls-m}\gamma_5\bs\frac{1}{\ls-m}
%\times\nonumber\\&\times&
\gamma^{\nu}\frac{1}{\ls+\ps-m}\gamma_5\bs\frac{1}{\ls+\ps-m}+\nonumber\\
&+&
\gamma^{\mu}\frac{1}{\ls-m}\gamma_5\bs\frac{1}{\ls-m}
%\times\nonumber\\&\times&
\gamma_5\bs\frac{1}{\ls-m}\gamma^{\nu}\frac{1}{\ls+\ps-m})
A_{\mu}(-p)A_{\nu}(p),
\label{ee}
\end{eqnarray}
%\end{widetext}
where $p$ is an external momentum.
Let us write
\be
S(l+p)=\frac{1}{\ls+\ps-m},
\ee
which can be expanded as
\be
S(l+p)=\sum_{i=0}^{\infty} \frac{1}{\ls-m}\left( -\ps \frac{1}{\ls-m}\right)^i = \sum_{i=0}^{\infty} S_i,
\ee
where
\be
S_i\equiv \frac{1}{\ls-m}\left( -\ps \frac{1}{\ls-m}\right)^i.
\ee
The expression of eq.(\ref{ee}) can then be written as
\bq
&&S_{AA}(p)=\frac{ie^2}{2}\int\frac{d^4l}{(2\pi)^4}\mbox{tr} \left\{ \gamma^\mu S_0 \gamma^\nu (S_0+S_1+S_2) \times \right. \nonumber \\
&& \left. \times \gamma_5\bs (S_0+S_1+S_2) \gamma_5\bs (S_0+S_1+S_2) + \right. \nonumber \\
&& \left. + \gamma^\mu S_0 \gamma_5 \bs S_0 \gamma^\nu (S_0+ S_1+S_2)\gamma_5 \bs (S_0+S_1+S_2) + \right. \nonumber \\
&& \left.+ \gamma^\mu S_0 \gamma_5 \bs S_0 \gamma_5 \bs S_0\gamma^\nu (S_0+S_1+S_2) \right\}
A_{\mu}(-p)A_{\nu}(p) + 
%\nonumber \\&&+ 
{\cal O}(p^3).
\eq
The contributions we are interested in are quadratic in $p$. These terms are finite by power counting and, thus, ambiguity-free. After a
straightforward calculation, in which we keep only the terms of second order in $p$ from the expression above (terms proportional to one $S_2$ or to two $S_1$), we obtain
\begin{eqnarray}
\label{aa}
S_{AA}=-\frac{e^2}{6m^2\pi^2}b_{\mu}F^{\mu\nu}b^{\lambda}F_{\lambda\nu}.
\end{eqnarray}
Indeed, we note that this  reproduces the form of the aether term obtained in \cite{aether} through a purely nonminimal interaction, while the numerical coefficient is naturally different.

Now, let us consider the ``mixed'' Feynman diagrams, involving both minimal and nonminimal couplings, and depicted at Fig.1.

\begin{figure}[ht]
\centerline{\includegraphics{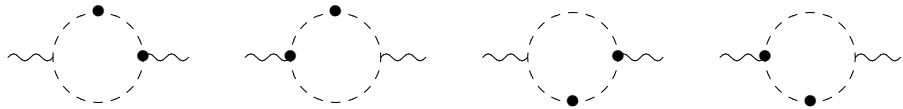}}
\caption{Contributions to the two-point function of the vector field.}
\end{figure}

Similarly to the calculations in \cite{MNP}, here we consider the Lorentz-breaking insertions $\gamma_5\bs$ introduced both in the vertices and in the propagators (both these insertions are denoted by the symbol $\bullet$).
The calculations do not essentially differ from those ones carried out in \cite{MNP}, since the loop integrals and traces are identically the same (actually, the only difference from those papers is related to the fact that in one of the vertices, the $A_{\mu}$ field is replaced by its ``dual'' $\epsilon_{\mu\nu\lambda\rho}b^{\nu}F^{\lambda\rho}$). As a result, we arrive at
\begin{eqnarray}
\label{Ef3}
S_{AF}&=&eg\int d^4x\;
k_{\rho}\epsilon^{\rho\nu\lambda\mu}
(\epsilon_{\nu\alpha\beta\gamma}F^{\alpha\beta}b^{\gamma} \partial_{\lambda}A_{\mu}+ %\nonumber\\&+&
A_{\nu}\epsilon_{\mu\kappa\eta\sigma}\partial_{\lambda}F^{\kappa\eta}b^\sigma),
\end{eqnarray}
where $k_{\rho}$ is a constant vector whose explicit form is
\begin{eqnarray}
\label{I1}
k_{\rho}=2i\int\,\frac{d^4l}{(2\pi)^4}\,\frac{b_{\rho}(l^2+3m^2)-4l_{\rho}(b\cdot l)}{(l^2-m^2)^3}.
\end{eqnarray}
It is well known (see f.e. \cite{ourLV1,ourLV2}) that the integral (\ref{I1}) is ambiguous, looking like $k_{\rho}=Cb_{\rho}$, while a finite constant $C$ crucially depends on the regularization prescription: within different procedures, it is equal to $\frac{1}{4\pi^2}$, $\frac{3}{8\pi^2}$, $\frac{3}{16\pi^2}$, zero, etc.

Multiplying the Levi-Civita symbols, we arrive at
\begin{eqnarray}
\label{af}
S_{AF}=-2Ceg b_{\mu}F^{\mu\nu}b^{\lambda}F_{\lambda\nu}.
\end{eqnarray}
So, we see that the ambiguity of the aether term arises from the ``mixed'' contribution involving both minimal and nonminimal couplings. Its form is similar to the usual ambiguity of the CFJ term (and of the higher-derivative terms \cite{MNP})  which essentially differs from the ambiguity of the aether term arisen from the purely nonminimal sector (\ref{s2d40},\ref{ff}). Therefore, we conclude that even the introduction of the minimal interaction (as it has been done in \cite{MCN}) does not allow to rule out the ambiguity of the aether term.

The complete result for the one-loop two-point function of the gauge field $A_{\mu}$ is a sum of (\ref{ff},\ref{aa},\ref{af}). It looks like
\begin{eqnarray}
\Gamma_2&=&S_{AA}+S_{AF}+S_{FF}=\nonumber\\
&=&(C_0g^2m^2-\frac{e^2}{6m^2\pi^2}-2Ceg)b_{\mu}F^{\mu\nu}b^{\lambda}F_{\lambda\nu}.
\end{eqnarray}
We conclude that this two-point function, first, displays the characteristic structure of the aether term \cite{aether}, and, second, is finite and ambiguous. Moreover, it involves two ambiguous dimensionless constants $C_0$ and $C$. We note that while the regularization where $C_0=0$ can be treated as the preferable since it allows to cancel the undesirable Proca term (see discussion in \cite{MCN}), there is no profound reason to prefer any value of $C$.

We would like to comment why we choose specifically the model of (\ref{mcn}). There are many ways to generate a CPT-even term in the gauge sector. For example, it can be generated by a chiral CPT-odd nonminimal coupling \cite{chiral-NM} or by a CPT-even nonminimal coupling \cite{NM-even}. If we consider the gauge sector, the CFJ term could radiatively induce the aether term at the two-loop order, or even at the one-loop order, if spontaneous gauge symmetry breaking takes place \cite{cleber}. However, {\bf just} as it occurs for the induction of the CFJ term, although the present model is very particular, it
is sufficient for our purposes, since it has an interesting peculiarity, that is, the finiteness accompanied by unavoidable ambiguity in the coefficient of the induced CPT-even term.

\section{Generation of the aether term for the massless model}
\label{sec:2}
In the previous section, we have considered the massive fermions case. However, it is instructive to consider also the massless fermions since in this case the calculations are much simpler. Just for comparison, we use two approaches.

\subsection{Functional calculation}

In the massless case, the fermion action is
given by
\begin{equation}
\Sigma_\psi = \int d^4x \,\, \bar \psi (i \ds - \tilde \As - \bs \gamma_5) \psi,
\end{equation}
which is the action (\ref{mcn}) with $m=0$, where we used the redefined gauge field
\begin{equation}
\tilde A_\mu= e A_\mu +g \varepsilon_{\mu \nu \alpha \beta} b^\nu F^{\alpha \beta}.
\label{redef}
\end{equation}
It is instructive to discuss first a nonperturbative (that is, exact both in $b$ and in the coupling constant) calculation of the induced
Lorentz violating terms. This calculation is similar to that one performed in \cite{Chung}.
By making the chiral transformation,
\begin{equation}
\psi \to e^{-i \gamma_5 b \cdot x}\psi \,\, , \,\, \bar \psi \to \bar \psi e^{-i \gamma_5 b \cdot x},
\end{equation}
we can eliminate the $b_\mu$ vector from the classical action. Nevertheless, repeating the arguments from \cite{Fujikawa}, one can show that, at the quantum level, the measure of the generating functional
acquires a factor given by the following Jacobian:
\begin{equation}
J[b_\mu,\tilde A_\mu]= \exp\left\{ -i\int d^4x \,\, (b \cdot x) {\cal A}[\tilde A_\mu](x) \right\},
\end{equation}
with
\begin{equation}
{\cal A}[\tilde A_\mu](x)=\frac {1}{16 \pi^2}\varepsilon^{\mu \nu \alpha \beta}\tilde F_{\mu \nu}\tilde F_{\alpha \beta},
\end{equation}
where $\tilde F_{\mu \nu}=\partial_\mu \tilde A_\nu - \partial_\nu \tilde A_\mu$.
We can write
\begin{eqnarray}
&&J[b_\mu,\tilde A_\mu]= \exp\left\{ -\frac{i}{4\pi^2}\int d^4x \, (b \cdot x)\epsilon^{\mu \nu \alpha \beta}
%\times\nonumber\\&\times&
\partial_\mu \tilde A_\nu \partial_\alpha \tilde A_\beta \right\},
\end{eqnarray}
which after an integration by parts turns out to be
\begin{equation}
J[b_\mu, \tilde A_\mu]= \exp\left\{ \frac{i}{4\pi^2}\int d^4x \,\, b_\mu\epsilon^{\mu \nu \alpha \beta}
\tilde A_\nu \partial_\alpha \tilde A_\beta \right\}.
\end{equation}
We see that after the chiral transformation the axial term disappears from the fermionic sector.
As a result the QED Lagrangian is obtained, together with a Jacobian which is taken into account when
quantum corrections are calculated.

We now can return to the usual gauge field and obtain the complete induced Lorentz violating terms. The induced
Lagrange density is given by
\begin{equation}
{\cal L}_{ind}= {\cal L}_b +{\cal L}_{bb} + {\cal L}_{bbb},
\end{equation}
with
\begin{equation}
{\cal L}_b= \frac{1}{4 \pi^2} e^2 b_\mu \varepsilon^{\mu \nu \alpha \beta} A_\nu \partial_\alpha A_\beta,
\end{equation}
\begin{eqnarray}
\label{measure}
&&{\cal L}_{bb}=\frac{1}{4 \pi^2} eg b_\mu\left(b^\theta \varepsilon^{\mu \nu \alpha \beta}
\varepsilon_{\beta \theta \delta \tau} A_\nu \partial_\alpha F^{\delta \tau}+
%\right. \nonumber\\&&+\left.
b^\rho \varepsilon^{\mu \nu \alpha \beta} \varepsilon_{\nu \rho \sigma \lambda} \partial_\alpha A_\beta F^{\sigma \lambda} \right)\nonumber \\
&&= \frac{1}{4 \pi^2} eg \left(-b^2 F_{\alpha \beta}F^{\alpha \beta}+4(b_\mu F^{\mu \alpha})^2\right)
\end{eqnarray}
and
\begin{eqnarray}
&&{\cal L}_{bbb}=\frac{1}{4\pi^2} g^2b_\mu b^\rho b^\theta F^{\sigma \lambda} \partial_\alpha F^{\delta \tau}
\varepsilon^{\mu \nu \alpha \beta} \varepsilon_{\nu \rho \sigma \lambda} \varepsilon_{\beta \theta \delta \tau} \nonumber \\
&&=-\frac{1}{4 \pi^2} g^2 \left(2b^2 b^\theta F^{\alpha \beta} \partial_\alpha F^{\delta \tau} \varepsilon_{\beta \theta \delta \tau}-
%\right.\nonumber\\&&-\left.
2b_\mu b^\alpha b^\theta F^{\mu \beta} \partial_\alpha F^{\delta \tau}\varepsilon_{\beta \theta \delta \tau}\right).
\end{eqnarray}

The three terms above give the total contribution to the photon sector arisen due to the axial term.
The first one is simply the induced Chern-Simons term which has been {\bf intensively} discussed in the last decade (see f.e. \cite{list,list1,list2}). The second one contains the aether term and the rescaled Maxwell term. The third one is composed by higher derivative terms whose different aspects have been considered in \cite{MNP,Mariz,Mariz1,MP}.

A comment on the coefficients obtained above is in order. The Fujikawa procedure for the calculation of the Jacobian indeed makes use of
a regularization procedure. An extension of this approach has been developed in \cite{Urrutia}, where a Jacobian dependent
on an undetermined parameter which must be adjusted to satisfy the Ward identity was obtained. Alternatively, as is was discussed in \cite{Chung}, there is a
freedom in the definition of the chiral current, considering all the properties it should satisfy, such that
\be
J_5^\mu= \bar \psi \gamma^\mu \gamma_5 \psi+c \varepsilon^{\mu \nu \alpha \beta}\tilde F_{\nu \alpha}\tilde A_\beta,
\ee
with $c$ an undetermined constant. Thus, all the coefficients obtained above are indeed ambiguous.

\subsection{The massless complete one-loop calculation}

We now carry out a one-loop analysis for the massless case. It is easy to do this with use of the modified gauge field $\tilde A_\mu$, since the complete one-loop amplitude can be written in terms of the vacuum polarization tensor of the modified massless QED, with only the axial term is added. In principle, one can
use each of many results obtained by a great number of regularization techniques. We use that one developed in \cite{scarp1}, since its
expression in function of surface terms allows us to discuss different possibilities and to identify the regularization dependence of the induced terms. The following result has been established in \cite{scarp1}:
\begin{equation}
T^{\mu \nu}=T^{\mu \nu}_0+T^{\mu \nu}_{b}+T^{\mu \nu}_{bb},
\end{equation}
with
\begin{eqnarray}
&&T^{\mu \nu}_0=\Pi(p^2)(p^\mu p^\nu-p^2 g^{\mu \nu}) -4 \alpha_1 g^{\mu \nu}
\nonumber \\&&
-\frac 43 \left[ \alpha_2(p^\mu p^\nu-p^2 g^{\mu \nu}) 
%\right. \nonumber \\&& \left. 
+ (2p^\mu p^\nu +p^2 g^{\mu \nu})(\alpha_3-2 \alpha_2)\right],
\label{T0}
\end{eqnarray}
\begin{equation}
\label{b1}
T^{\mu \nu}_b=-4 i  \alpha_2 p_\alpha b_\beta \epsilon^{\mu \nu \alpha \beta}
\label{T1}
\end{equation}
and
\begin{equation}
T^{\mu \nu}_{bb}=-4\left\{ \left(b^2g^{\mu \nu} +2 b^\mu b^\nu \right)(\alpha_3-2\alpha_2) \right\},
\label{T2}
\end{equation}
where the coefficients $\alpha_1,\alpha_2,\alpha_3$ are introduced in \cite{scarp1}. Their explicit form is
\begin{eqnarray}
&&\alpha_1 g_{\mu\nu}=\int\frac{d^4k}{(2\pi)^4}\frac{\partial}{\partial k^{\mu}}\frac{k_{\nu}}{k^2-\lambda^2};\nonumber\\
&&\alpha_2 g_{\mu\nu}=\int\frac{d^4k}{(2\pi)^4}\frac{\partial}{\partial k^{\mu}}\frac{k_{\nu}}{(k^2-\lambda^2)^2};\nonumber\\
&&\alpha_3g_{\{\mu \nu}g_{\alpha \beta\}}=\int^\Lambda \frac{d^4k}{(2\pi)^4}\frac{\partial}{\partial k^\beta}
\left[ \frac{4k_\mu k_\nu k_\alpha}{(k^2-\lambda^2)^3} \right]+
%\nonumber \\&&+
\alpha_2g_{\{\mu \nu}g_{\alpha \beta\}},
\end{eqnarray}
where $g_{\{\mu \nu}g_{\alpha \beta\}}$ is a symmetrized product of two Minkowski metrics and $\lambda$ is a mass scale.

Let us argue that the term relevant for our purpose is that one linear in $b^{\mu}$ (\ref{b1}). In principle, the complete one-loop
photon self-energy for the present model can involve fourth order in $b^{\mu}$, because of the quadratic part $T^{\mu \nu}_{bb}$ and
the two additional $b^{\mu}$ arising from the nonminimal vertices. However, a simple observation of the equations (\ref{T0}) and (\ref{T2})
shows that the condition of transversality of $T^{\mu \nu}$ imposes the surface terms to respect the relations $\alpha_1=0$ and
$\alpha_3=2\alpha_2$. These conditions imply that $T^{\mu \nu}_{bb}=0$. So the higher order term is of the order $b^3$.

Let us also discuss
the contribution to the Lorentz violating part coming from $T^{\mu \nu}_0$. In \cite{MCN}, it was shown that, since we are interested
in the limit $p^2 \to 0$, the contributions to the CPT odd and even terms coming from $T^{\mu \nu}_0$ have the fermion mass in their
coefficients. So, for this massless case, they do not contribute. We are then left with $T^{\mu \nu}_b$.

We have now a simple task, if we use the redefined field $\tilde A_\mu$ of equation (\ref{redef}). We can write the induced term as
\begin{equation}
{\cal L}_{LV}=-2i \alpha_2 b_\mu \varepsilon^{\mu \nu \alpha \beta} \tilde A_\nu \partial_\alpha \tilde A_\beta.
\end{equation}
With a simple substitution in terms of the $A_\mu$ field we obtain the complete one loop Lorentz violating induced term:
\begin{eqnarray}
\label{complete}
&&{\cal L}_{LV}= -2i \alpha_2 e^2 b_\mu \varepsilon^{\mu \nu \alpha \beta} A_\nu \partial_\alpha A_\beta
%\nonumber \\&&
-2i \alpha_2 eg \left(-b^2 F_{\alpha \beta}F^{\alpha \beta}+4(b_\mu F^{\mu \alpha})^2\right) \nonumber \\
&& -2i \alpha_2 g^2 \left(2b^2 b^\theta F^{\alpha \beta} \partial_\alpha F^{\delta \tau} \varepsilon_{\beta \theta \delta \tau}
%\right.\nonumber\\&&\left.
-2b_\mu b^\alpha b^\theta F^{\mu \beta} \partial_\alpha F^{\delta \tau}\varepsilon_{\beta \theta \delta \tau}\right).
\end{eqnarray}

It is worth
to note that all these terms are proportional to the factor $\alpha_2$ whose value cannot be fixed by gauge invariance
reasons, the only possible restriction involving $\alpha_2$ is $\alpha_3=2\alpha_2$. This fact reveals an unavoidable ambiguity of the coefficients of the Lorentz
violating induced terms. This differs from the case studied in \cite{MCN}, where the axial term was absent, and the coefficient
of the induced terms, $\alpha_1$, should be fixed to be equal to zero in order to preserve the transversality of the vacuum polarization tensor of the traditional QED sector. In particular, we conclude that the aether-like contribution is essentially ambiguous.

\section{Summary}

Now, let us discuss our results. We have considered the perturbative generation of the aether-like term in the extended Lorentz-breaking QED whose action involves both couplings, the minimal one and the nonminimal one \cite{MCN} and, besides, an axial term in the fermionic sector. We have found that this term is exactly the same one considered in \cite{aether}, being gauge invariant and UV finite despite the superficial logarithmic divergence of the corresponding contribution.  We note that this term emerges even at $g=0$, that is, in the case of the pure Lorentz-breaking QED without the nonminimal interaction. It is worth mentioning that just in this case, the aether term is not ambiguous. We note that, besides the ambiguity of the aether term discussed earlier in \cite{aether,MCN} and characterized by the coefficient $C_0$, a new ambiguity described by the coefficient $C$, which is identically the same as that one accompanying the CFJ term in the usual Lorentz-breaking QED \cite{ourLV1,ourLV2}, also arises. This shows that the ABJ anomaly which is known to be responsible for the ambiguity of the CFJ term \cite{JackAmb} can be naturally promoted to a wide class of new terms. One must note, however, that this ambiguity disappears if we switch off the nonminimal interaction. Therefore, it does not arise in the usual extended QED, although the aether-like term arises also in this case. Moreover, differently from the ambiguity of the aether term considered in \cite{aether,MCN}, the new ambiguity cannot be removed via the choice of a gauge-preserving regularization. Also, we note that the functional integral approach developed in \cite{Chung} and applied here for the massless fermions, can be naturally generalized for the massive fermions case, whereas this calculation seems to be very complicated from the technical viewpoint. We are planning to do it in a forthcoming paper.

{\bf Acknowledgements.} This work was partially supported by Conselho
Nacional de Desenvolvimento Cient\'{\i}fico e Tecnol\'{o}gico (CNPq). The work by A. Yu. P. has been supported by the
CNPq project No. 303438/2012-6.

\end{document}